\documentclass[letterpaper, 10 pt, conference]{ieeeconf}  

\IEEEoverridecommandlockouts                              

\usepackage[utf8]{inputenc}
\usepackage[english]{babel}
\usepackage{graphics} 
\usepackage{graphicx}
\usepackage[utf8]{inputenc}
\usepackage{times} 
\usepackage{amsmath} 
\usepackage{xcolor}

\usepackage{amsthm} 
\usepackage{amssymb}  
\usepackage{multirow}
\usepackage{textcomp}
\usepackage{subfigure}
\usepackage[style=ieee,sorting=none,maxnames=3,doi=false,url=false,isbn=false]{biblatex}
\AtEveryBibitem{\clearfield{issn}}
\AtEveryCitekey{\clearfield{issn}}
\AtEveryBibitem{\clearfield{arxivId}}

\title{\LARGE \bf
Closed-loop Control Design and Motor Allocation for a Lower-limb Cable-driven Exoskeleton: A Switched Systems Approach
}

\author{Chen-Hao Chang$^{1}$, Jonathan Casas$^{1}$, Victor H. Duenas$^{1}$
\thanks{$^{1}$Department of Mechanical and Aerospace Engineering, Syracuse University, Syracuse NY 13244, USA. Email: {\tt\small \{cchang21, jacasasb,  vhduenas\}@syr.edu}}
}

\theoremstyle{plain}
\newtheorem{theorem}{Theorem}

\newtheorem{prop}{Property}

\theoremstyle{remark}

\begin{document}
\maketitle
\thispagestyle{empty}
\pagestyle{empty}
\begin{abstract}
Powered lower-limb exoskeletons provide assistive torques to coordinate limb motion during walking in individuals with movement disorders. Advances in sensing and actuation have improved the wearability and portability of state-of-the-art exoskeletons for walking. Cable-driven exoskeletons offload the actuators away from the user, thus rendering light-weight devices to facilitate locomotion training. However, cable-driven mechanisms experience a slacking behavior if tension is not accurately controlled. Moreover, counteracting forces can arise between the agonist and antagonist motors yielding undesired joint motion. In this paper, the strategy is to develop two control layers to improve the performance of a cable-driven exoskeleton. First, a joint tracking controller is designed using a high-gain robust approach to track desired knee and hip trajectories. Second, a motor synchronization objective is developed to mitigate the effects of cable slacking for a pair of electric motors that actuate each joint. A sliding-mode robust controller is designed for the motor synchronization objective. A Lyapunov-based stability analysis is developed to  guarantee a uniformly  ultimately  bounded  result  for  joint  tracking  and exponential tracking for the motor synchronization objective. Moreover, an average dwell time analysis provides a bound on the number of motor switches when allocating the control between motors that actuate each joint. An experimental result with an able-bodied individual illustrates the feasibility of the developed control methods.
\end{abstract}


\section{Introduction}

Regaining mobility is a top priority for individuals with movement disorders, whose locomotion is affected by muscle weakness and reduced leg coordination. Powered lower-limb exoskeletons aim to restore and improve function of people with spinal cord injury and post-stroke. Powered exoskeletons vary in their design, wearability, sensing, and actuation. Existing powered exoskeletons use electrical motors \cite{Rodriguez-Fernandez2021}, hydraulic \cite{Chang2017b} and pneumatic actuators \cite{Shorter2011,Chin2009}. The actuators used to outfit the exoskeleton influence the magnitude and response of the inputs applied to the body and thus the human-robot interaction. Particularly, traditional exoskeletons use frames to affix actuators to the human body, however this approach results in increased system's inertia and metabolic costs of walking~\cite{Sanjuan2020}. Therefore, the design of portable and lighter exoskeleton devices has been the focus of recent research to achieve transparent motion and enable human volition~\cite{Lv.Zhu.ea2018}. 

Cable-driven transmission mechanisms offload actuators away from the human to reduce the weight imposed on the body and thus reduces the burden on the user-side. Such mechanisms have been widely applied for orthotic devices (e.g., lower-limb~\cite{Bryan2020,Asbeck2015}, upper-limb~\cite{Noda2014,Lenzi2011}, hand orthoses~\cite{Chiri2009}) to provide torques about the joints to assist or augment human function.
Since cables cannot transmit compression forces, at least $l+1$ cables are required to control $l$ DOFs to provide an agonist-antagonist movement~\cite{Verhoeven2003}. Despite the benefits of cable-driven mechanisms, two critical issues arise: (1) cables experience a slack behavior if the tension is not accurately controlled, thus affecting the response time; (2) undesired agonist-antagonist coordination may produce counteracting torques about a joint. Hence, it is essential to develop effective control strategies to ensure coordinated or synchronized motion in a multi-joint system.       

Multiple strategies have been developed to mitigate the undesirable slacking behavior and coordination issues in cable-driven systems. A mechanical approach with closed pulley transmission (one motor with two cables) was used in~\cite{Wu2020} to achieve bi-directional joint movement. Alternatively, a motor synchronization approach was developed in~\cite{Bryan2020}, where one motor controlled the joint torque, and its antagonist motor was synchronized using a closed-loop proportional controller. The control designs in~\cite{Okur2015,Lu2017} combined the robot and actuator dynamics and provided proofs of stability. In this paper, the strategy is to develop controllers for a cable-driven exoskeleton by segregating the control into layers: the motor layer to mitigate the slacking behavior in cables that actuate joints, and the joint layer to achieve the tracking objective. This layered approach is developed to improve the scalability and prototyping of gait controllers. 



In a typical multi-layer control system, the higher layer usually generates the gait pattern to be tracked~\cite{Ouyang2021} or force fields~\cite{Mao2011}. The lower layer focuses on allocating control among actuators and generate control inputs. In this paper, the control allocation is addressed using switching control, which requires a switched-based stability analysis. Up to the best knowledge of the authors, the existing literature on cable-driven exoskeletons has not exploited switched systems control and analysis to improve tracking performance. Particularly, an average dwell time analysis as in \cite{Liberzon2003} is introduced to bound the number of switches (i.e., control allocation switches) within a time period to prevent counteracting forces from being applied about a joint by a pair of electric motors. 

In this paper, the control design to actuate leg joints using cable-driven mechanisms is segregated into layers. First, the joint control layer generates the control input to track the desired joint trajectory. A high-gain robust kinematic tracking controller is designed for the hip and knee joints. Second, the motor control layer includes a pair of electric motors to provide bidirectional motion about a joint. For any joint actuated by a pair of electric motors, the motor that provides cable tension to actuate the joint is called the lead motor. The other motor is called the follower motor. Therefore, the lead motor receives the joint kinematic control input to actuate the joint in the desired direction (e.g., joint flexion). Meanwhile, a closed-loop robust sliding-based controller is developed for the follower motor to ensure synchronization with the lead motor (i.e., mitigate the potential slacking behavior when not actuating the joint). Once the desired joint direction reverses (e.g., joint extension), the previous lead motor becomes the follower motor, and vice versa. A Lyapunov-based stability analysis is provided to guarantee uniformly ultimately bounded result for joint tracking and exponential tracking for the motor synchronization objective (i.e., prevent cable slacking). An average dwell time analysis provides a bound on the number of motor switches when allocating the control between motors that actuate each joint. An experimental result with an able-bodied individual is presented to illustrate the performance of developed controllers.

\section{Human-Exoskeleton dynamic model}
A motorized exoskeleton with a user can be modeled as a four-link bipedal walking system in the sagittal plane as follows
\begin{equation}
M(q)\ddot{q}+C(q,\dot{q})\dot{q}+G(q)+P(q,\dot{q})+d(t)=\tau_{e}(q,\dot{q},t),
\label{eq:exo dynamic}
\end{equation}
where $q:\mathbb{R}_{\geq t_{0}}\rightarrow\mathbb{R}^{4}$
denotes the measurable hip and knee joint angles on both sides, $\dot{q},\ddot{q}:\mathbb{R}_{\geq t_{0}}\rightarrow\mathbb{R}^{4}$ denote the measurable joint angular velocities and unmeasurable joint angular accelerations, respectively, where $t_{0}\in\mathbb{R}_{>0}$ is denoted as initial time; $M:\mathbb{R}^{4}\rightarrow\mathbb{R}_{>0}^{4\times4}$ denotes the combined human-exoskeleton inertia; $C:\mathbb{R}^{4}\times\mathbb{R}^{4}\rightarrow\mathbb{R}^{4\times4}$ and $G:\mathbb{R}^{4}\rightarrow\mathbb{R}^{4}$ denote centripetal-Coriolis and gravitational effects, respectively; $P:\mathbb{R}^{4}\times\mathbb{R}^{4}\rightarrow\mathbb{R}^{4}$ denotes the damping and viscoelastic effects; and $d:\mathbb{R}_{\geq t_{0}}\rightarrow\mathbb{R}^{4}$ denotes unmodeled terms and disturbances. The torque produced by the electrical motors are expressed as $\tau_{e}:\mathbb{R}^{4}\times\mathbb{R}^{4}\times\mathbb{R}_{\geq t_{0}}\rightarrow\mathbb{R}^{4}$. A body-weight support system is utilized to support and stabilize the trunk (i.e., trunk dynamics are not considered).

\subsection{Cable-driven Actuator System}
The lower-limb exoskeleton is actuated by electric motors utilizing customized cable-driven mechanisms. Electric motors can be segregated into extension (ex) and flexion (fl) actuator groups with respect to each joint. Forces are transmitted to each joint via Bowden cables. A pair of motors control extension and flexion for each joint. Four pairs of motors are used to control bilaterally the hip and knee joints. One of the motors in a pair, called the lead motor, receives the control input of the joint tracking controller to achieve the desired motion (e.g., flexion or extension). The second motor in a pair, called the follower motor, rotates in the opposite direction to the lead motor. The follower motor receives the input of a synchronization controller, which prevents force tension conflicts with the lead motor, mitigate cable slack, and provides fast force response.

\begin{figure}
    \centering
    \includegraphics[width=1\columnwidth]{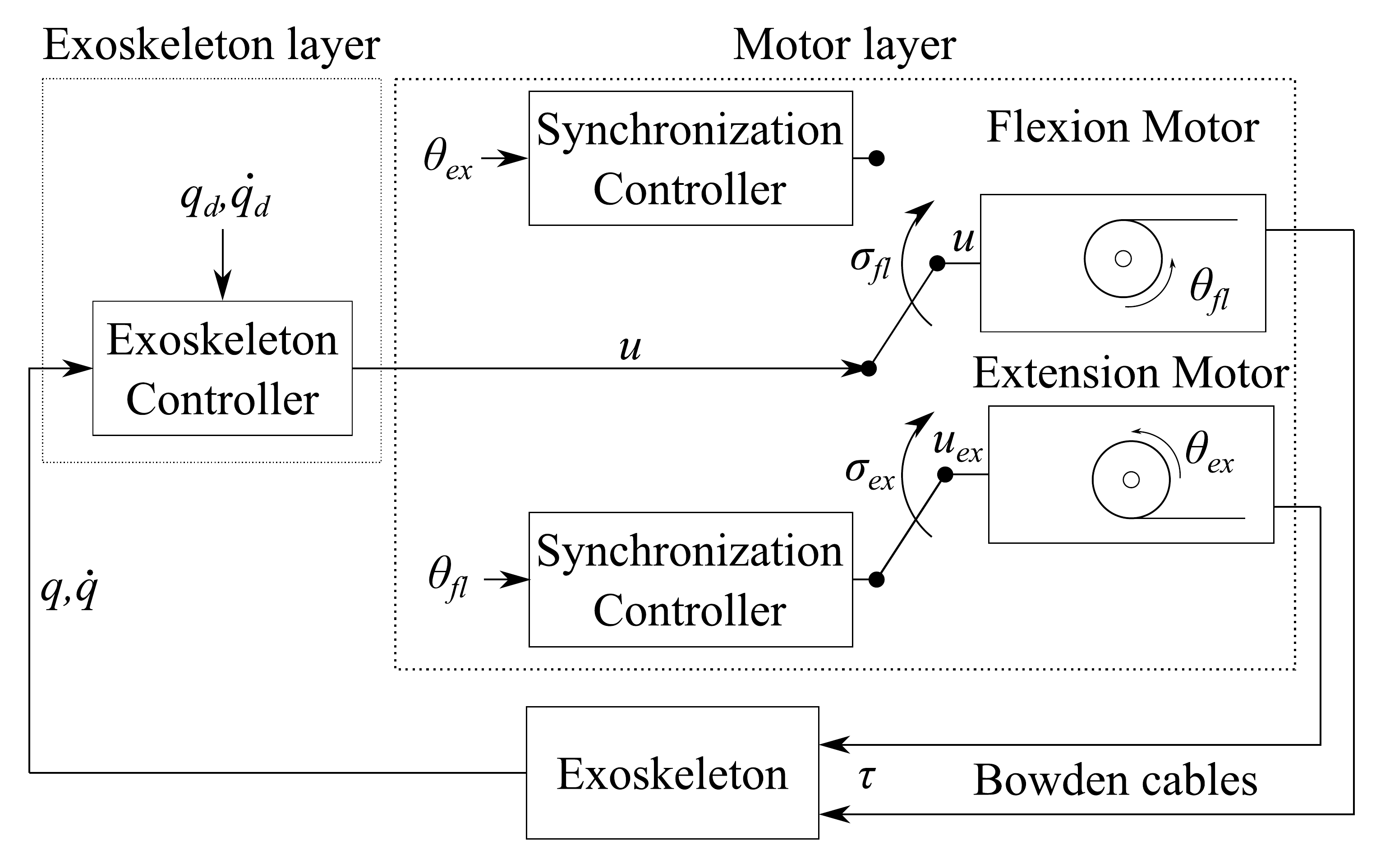}
    \caption{Schematic representation for the multi-layer control system (illustrated for flexion motion). The joint control layer sends a flexion control input ($u>0$) to the motor control layer. The switching signal is triggered as $\sigma_{fl}=1$ to allocate the control input to the flexion motor. The extension motor (follower motor in this case, i.e., $\sigma_{ex}=0$ ) is allocated to follow the angle of the flexion motor (lead motor) commanded by the motor synchronization controller ($u_{ex}$).}
    \label{fig:high/low}
\end{figure}

The torque produced by the motors to actuate the exoskeleton is defined as
\begin{align}
    \tau_{e}(q,\dot{q},t) \triangleq & B_\sigma(q,\dot{q}) u(t),\label{eq:tau motor}
    \\B_\sigma(q,\dot{q})\triangleq & \sum_{n=1}^8B_{n}\sigma_{n}(t),\label{eq:B sigma}
\end{align}
where the subscript $n\in\{1,2,...,8\}\subset \mathbb{N}$
denotes the motor index, and let $\mathcal{E},\mathcal{F}$ the set of all extension and flexion motors, respectively. The control input $u:\mathbb{R}_{\geq t_{0}}\rightarrow\mathbb{R}^4$ is designed in Section \ref{sec: Control Development}, where positive and negative control inputs refer to flexion and extension movements, respectively. The unknown individual motor control effectiveness is denoted as $B_{n}:\mathbb{R}_{>0}^{4\times4}$, $\forall n\in\mathcal{E}\cup\mathcal{F}$. A piecewise constant switching signal for each motor $\sigma_{n}(t)\in\{0,1\}$ determines if the motor is applying exoskeleton control input. 
The subscript $\sigma\in\mathcal{S}$, such that $\mathcal{S}=\{1,2,3,\dots,i\}$ denotes the \textit{$i^{th}$} possible lead motors combination. The lumped switched motor control effectiveness is denoted as $B_\sigma$. A schematic representation of the time-varying control system is presented in Figure \ref{fig:high/low}.
The following properties are exploited in the subsequent control design and stability analysis. 
\begin{prop}
\label{prop:bounded_M}
 $c_m\left\Vert\xi\right\Vert^2\leq \xi^T M(q) \xi\leq c_M\left\Vert\xi\right\Vert^2$, $\forall \xi\in\mathbb{R}^{4}$, where $c_m$ and $c_M$ are known positive constants.
\end{prop}


\begin{prop}
\label{prop:bounded_C}
$\left\Vert C(q,\dot{q})\right\Vert\leq c_c\left\Vert\dot{q}\right\Vert$, where $c_c$ is a known positive constant.
\end{prop}

\begin{prop}
\label{prop:bounded_G}
$\left\Vert G(q)\right\Vert\leq c_g$, where $c_g$ is a known positive constant.
\end{prop}

\begin{prop}
\label{prop:bounded_P}
$\left\Vert P(q,\dot{q})\right\Vert\leq c_{p1}+c_{p2}\left\Vert\dot{q}\right\Vert$, where $c_{p1}$ and $c_{p2}$ are known positive constants.
\end{prop}

\begin{prop}
\label{prop:bounded_d}
$\left\Vert d(t)\right\Vert\leq d_{exo}$, where $d_{exo}$ is a known positive constant.
\end{prop}

\begin{prop}
\label{prop:bounded_B}
 $\underline{B}\left\Vert\xi\right\Vert^2\leq \xi^T B_{\sigma} \xi\leq \overline{B}\left\Vert\xi\right\Vert^2$, $\forall \xi\in\mathbb{R}^{4}$, $\forall\sigma\in\mathcal{S}$, where $\underline{B}$ and $\overline{B}$ are known positive constants.
\end{prop}

\begin{prop}
\label{prop:skew_symmetry}
The skew-symmetry $\xi^T\left(\frac{1}{2}\dot{M}-C\right)\xi=0 $, $\forall \xi\in\mathbb{R}^{4}$ \cite{Lewis2004}.
\end{prop}
\subsection{Follower Motor Model}
Each motor system in the exoskeleton includes an electric motor, a gearbox and a pulley. The motor system dynamics are modeled to develop the motor synchronization controller as follows \cite{Lewis2004}
\begin{equation}
J_{n}\ddot{\theta}_n+D_n\dot{\theta}_n+d_n(t)=B_n u_n\bar{\sigma}_n(t),
\label{eq:motor dynamic}
\end{equation}
where $\theta_{n},\dot{\theta}_n,\ddot{\theta}_n$ denotes measurable motor's angle, angular velocity, and unmeasurable angular acceleration  of the $n^{th}$ motor system. The inertia $J_{n}\in\mathbb{R}_{>0}$, damping constant $D_n\in\mathbb{R}$, and disturbances with friction and any unmodeled terms $d_n\in\mathbb{R}$. The control input $u_n\in\mathbb{R}$ corresponding to the $n^\text{th}$ motor is designed in Section \ref{sec: Control Development}. A piecewise constant switching signal $\bar{\sigma}_n$ that determines if the motor is applying motor synchronization control input. The switching signals $\sigma_n$ and $\bar{\sigma}_n$ are opposite to each other, i.e., if the $n^\text{th}$ motor at time $t$ is performing motor synchronization, then $\bar{\sigma}_n(t)=1$, $\sigma_n(t)=0$, and vice versa. 
The following properties are exploited in the subsequent control design and stability analysis. 
\begin{prop}
\label{prop:motor_JDd}
$J_{n}, D_n, d_n(t)$ satisfy the inequalities $c_j\leq J_{n}\leq c_J$, $c_d\leq D_n\leq c_D$, $c_{de}\leq d_n(t)\leq c_{De}$, $\forall n\in\mathcal{E}\cup\mathcal{F}$, where $c_j$, $c_J$, $c_d$, $c_D$, $c_{de}$, and $c_{De}$ are known positive constants.
\end{prop}

\begin{prop}
\label{prop:motor bounded_B}
 $\underline{B}_n\left\Vert\xi\right\Vert^2\leq \xi^T B_{n} \xi\leq \overline{B}_n\left\Vert\xi\right\Vert^2$, $\forall \xi\in\mathbb{R}^4$, $\forall n\in\mathcal{E}\cup\mathcal{F}$ , where $\underline{B}_n$ and $\overline{B}_n$ are known positive constants.
\end{prop}
\section{Control Development} \label{sec: Control Development}
\theoremstyle{plain}
The control objectives of this paper are twofold. First, a high-gain tracking controller is developed to track desired hip and knee joint angles. Second, a robust sliding-mode controller is developed to achieve the synchronization objective for the follower motor.
\subsection{Joint Kinematic Control}
The measurable joint position trajectory tracking error $\xi:\mathbb{R}_{\ge t_{0}}\rightarrow\mathbb{R}^{4}$ and filtered tracking error $\eta:\mathbb{R}_{\ge t_{0}}\rightarrow\mathbb{R}^{4}$ are defined as
\begin{align}
    &\xi(t)\triangleq q_{d}(t)-q(t),\label{eq:high e}\\
    &\eta(t)\triangleq\dot{\xi}(t)+\alpha \xi(t),\label{eq:high r}
\end{align}
where $\alpha\in\mathbb{R}$ is a selectable positive control gain and $q_d(t),\dot{q}_d(t),\ddot{q}_d(t):\mathbb{R}_{\ge t_{0}}\rightarrow\mathbb{R}^4$ are bounded desired trajectories. The open-loop error system is obtained by taking the time derivative of (\ref{eq:high r}), pre-multiplying $M$, substituting (\ref{eq:high e}), and performing algebraic manipulation as
\begin{equation}
    M\dot{\eta}=\chi-C\eta-B_\sigma u-\xi,
    \label{eq:high open-loop}
\end{equation}
where the auxiliary signal $\chi:\mathbb{R}_{\geq t_{0}}\rightarrow\mathbb{R}^4$ is defined as
\begin{equation}
    \chi\triangleq M\left(\ddot{q}_d+\alpha \dot{\xi}\right)+C\left(\dot{q}_d+\alpha \xi\right)+G+P+d.
    \label{eq:high chi}
\end{equation}
Using Properties \ref{prop:bounded_M}-\ref{prop:bounded_d}, the auxiliary signal can be upper bounded as 
\begin{equation}
    \left\Vert \chi\right\Vert \leq \rho\left(\Vert z_{1} \Vert\right),
    \label{eq:high chi upper bound}
\end{equation}
where $z_{1}\triangleq\left[\begin{array}{cc}
\xi^T & \eta^T\end{array}\right]^{T}:\mathbb{R}_{\geq t_{0}}\rightarrow\mathbb{R}^{8}$, $\rho\left(\Vert z_{1} \Vert\right)\triangleq\rho_1+\rho_2\Vert z_{1} \Vert+\rho_3\Vert z_{1} \Vert^2$, and $\rho_1,\rho_2,\rho_3\in\mathbb{R}_{\geq 0}$. The controller is designed as
\begin{equation}
    u=k_1 \eta+\frac{1}{\epsilon}\rho^2\left(\Vert z_{1} \Vert\right)\eta,
    \label{eq:high u}
\end{equation}
where $k_1, \epsilon\in\mathbb{R}_{>0}$ are positive selectable gains. The close-loop error system can be obtained by substituting (\ref{eq:high u}) into (\ref{eq:high open-loop}) as
\begin{equation}
    M\dot{\eta}=\chi-C\eta-\xi-B_\sigma\left(k_1 \eta+\frac{1}{\epsilon}\rho^2\left(\Vert z_{1} \Vert\right)\eta\right).
    \label{eq:high closed-loop}
\end{equation}
\subsection{Motor Synchronization Control}

Let $\varrho=\{ex,fl\}$ denote the pair of motors acting on a joint, where, $ex$ and $fl$ stand for the extension and flexion motors, respectively.  The measurable position tracking error $e:\mathbb{R}_{\ge t_{0}}\rightarrow\mathbb{R}$ and filtered tracking error $r:\mathbb{R}_{\ge t_{0}}\rightarrow\mathbb{R}$ are defined as 
\begin{align}
    &e(t)\triangleq \theta_{fl}(t)-\theta_{ex}(t),\label{eq:low motor e}\\
    &r(t)\triangleq\dot{e}(t)+\beta e(t),\label{eq:low motor r}
\end{align}
where $\beta \in\mathbb{R}_{>0}$ is a selectable positive control gain. The angles for flexion and extension motor are denoted as $\theta_{fl}(t)$ and $\theta_{ex}(t)$, respectively. The controllers for extension and flexion motor can be designed separately as follows.

\subsubsection{Extension motor (ex)}
The synchronization control for the extension motor is designed to track the flexion motor angle $\theta_{fl}$. Taking derivative of the filtered tracking error (\ref{eq:low motor r}), pre-multiplying by $J_{ex}$, substituting (\ref{eq:low motor e}), and then performing algebraic manipulation yields
\begin{equation}
    J_{ex}\dot{r}=\chi_{ex}-B_{ex} u_{ex}-e,
    \label{eq:low Ex open-loop}
\end{equation}
where the auxiliary signal $\chi_{ex}:\mathbb{R}\times\mathbb{R}_{\geq t_{0}}\rightarrow\mathbb{R}$ is defined as
\begin{equation}
    \chi_{ex}=J_{ex}\left(\ddot\theta_{fl}+\beta\dot{e}\right)+D_{ex}\left(\dot\theta_{fl}-\dot{e}\right)+d_{ex}+e.
    \label{eq:low Ex chi}
\end{equation}
Based on (\ref{eq:low Ex open-loop}) and the stability analysis in Section \ref{sec: stability}, the controller is designed as
\begin{equation}
    u_{ex}=k_{2} r+\left(k_{3}+k_{4}\left\Vert z_{2}\right\Vert\right)\text{sgn}(r),
    \label{eq:low Ex u}
\end{equation}
where $z_{2}\triangleq\left[\begin{array}{cc}
e & r\end{array}\right]^{T}:\mathbb{R}_{\geq t_{0}}\rightarrow\mathbb{R}^{2}$, and $k_{2}, k_{3}, k_{4}\in\mathbb{R}_{>0}$ are positive selectable gains. 

\subsubsection{Flexion motor (fl)}
The synchronization control objective for the flexion motor is to track the extension motor angle $\theta_{ex}$. Utilizing a similar process as for the control development of the extension motor yields
\begin{equation}
    J_{fl}\dot{r}=\chi_{fl}+B_{fl} u_{fl}-e,
    \label{eq:low Fl open-loop}
\end{equation}
where the auxiliary signal $\chi_{fl}:\mathbb{R}\times\mathbb{R}_{\geq t_{0}}\rightarrow\mathbb{R}$ is defined as
\begin{equation}
    \chi_{fl}=-J_{fl}\left(\ddot\theta_{ex}-\beta\dot{e}\right)-D_{fl}\left(\dot\theta_{ex}-\dot{e}\right)-d_{fl}+e.
    \label{eq:low Fl chi}
\end{equation}
Based on (\ref{eq:low Fl open-loop}) and the stability analysis in Section \ref{sec: stability}, the controller can be designed as

\begin{equation}
    u_{fl}=-k_{2} r-\left(k_{3}+k_{4}\left\Vert z_{2}\right\Vert\right)\text{sgn}(r).
    \label{eq:low Fl u}
\end{equation}

Given the open-loop dynamics in (\ref{eq:low Ex open-loop}), (\ref{eq:low Fl open-loop}) and control design in (\ref{eq:low Ex u}), (\ref{eq:low Fl u}), the closed-loop error dynamics can be compactly expressed as 
\begin{equation}
    J_\varrho\dot{r}=\chi_\varrho-e-B_\varrho\left(k_{2} r+\left(k_{3}+k_{4}\left\Vert z_2\right\Vert\right)\text{sgn}(r)\right).
    \label{eq:low closed-loop}
\end{equation}

Similarly, after leveraging the Property \ref{prop:motor_JDd}, the auxiliary signals in (\ref{eq:low Ex chi}) and (\ref{eq:low Fl chi}) can be upper bounded as
\begin{equation}
    \left\Vert \chi_\varrho\right\Vert \leq c_{1}+c_{2}\left\Vert z_{2}\right\Vert,
    \label{eq:low chi upper bound}
\end{equation}
where $c_{1},c_{2}\in\mathbb{R}_{>0}$ are positive bounding constants.
From Theorem \ref{th:theorem1}, the lead motor's control input $u$, joint angle, angular velocity, and acceleration are bounded, which implies that the desired synchronization trajectories $\theta_{fl}$ for the extension motor and $\theta_{ex}$ for the flexion motor are bounded. 

\section{Stability Analysis}
\label{sec: stability}
The following theorems examine the stability for the developed joint tracking and motor synchronization controllers. Theorem \ref{th:theorem1} demonstrates global uniformly ultimately bounded (GUUB) for the joint tracking errors. Theorem \ref{th:theorem2} shows exponential tracking of motor synchronization errors across multiple switching time period.  Theorem \ref{th:theorem3} uses an average dwell time analysis to provide a bound on the number of motor switches and ensures exponential tracking under fast switching.
\begin{theorem}
\label{th:theorem1}
Given the closed-loop error system in (\ref{eq:high closed-loop}), the controller in (\ref{eq:high u}) ensures globally uniformly ultimately bounded (GUUB) in the sense that
\begin{equation}
    V\leq V(0)e^{-\delta t}+\frac{\epsilon}{\delta}\left(1-e^{-\delta t}\right),
    \label{eq:high GUUB result}
\end{equation}
where $\delta=\frac{1}{b}\text{min}\{\alpha,\underline{B} k_{1}\}$.

\begin{proof}
Defining a nonnegative, continuously differentiable Lyapunov function  $V:\mathbb{R}^4\times\mathbb{R}^4\times\mathbb{R}_{\geq t_{0}}\rightarrow\mathbb{R}$ as
\begin{equation}
    V=\frac{1}{2}\xi^T\xi+\frac{1}{2}M\eta^T\eta,
    \label{eq:high V}
\end{equation}
which satisfies the following inequalities
\begin{equation}
    a\left\| z_{1}\right\|^2\le V(z_{1},t)\le b\left\| z_{1}\right\|^2,
    \label{eq:high V bounded}
\end{equation}
where $a=\min\left(\frac{1}{2},\frac{1}{2}c_m\right), b=\max\left(\frac{1}{2},\frac{1}{2}c_M\right)$ are positive known constants. After substituting for (\ref{eq:high closed-loop}) and canceling common terms, the time derivative of (\ref{eq:high V}) can be expressed as
\begin{multline}
    \dot{V}=-\alpha \xi^T\xi+\frac{1}{2}\dot{M}\eta^T\eta+ \eta^T\Bigg[\chi-C\eta
    \\-B_\sigma\left(k_1 \eta+\frac{1}{\epsilon}\rho^2\left(\Vert z_{1} \Vert\right)\eta\right)\Bigg].
    \label{eq:high V_dot}
\end{multline}
With the condition in (\ref{eq:high chi upper bound}), Property (\ref{prop:bounded_B}) and Property (\ref{prop:skew_symmetry}), (\ref{eq:high V_dot}) can be upper bounded as
\begin{equation}
    \dot{V}\leq-\alpha \xi^T\xi- \underline{B} k_{1} \eta^T\eta + \eta^T|\rho| \left[1-\frac{\underline{B} \eta|\rho|}{\epsilon}\right].
    \label{eq:high V_dot bounded}
\end{equation}
The inequality can be further upper bounded as
\begin{align}
    \dot{V}&\leq-\alpha \xi^T\xi- \underline{B} k_{1} \eta^T\eta + \epsilon,
    \\&\leq-\delta V + \epsilon.
    \label{eq:high V_dot final}
\end{align}
Based on (\ref{eq:high V bounded}) and (\ref{eq:high V_dot final}), the GUUB result in (\ref{eq:high GUUB result}) can be obtained. Since $V\in\mathcal{L}_\infty$, $\xi,\eta\in\mathcal{L}_\infty$ (i.e., $z_1\in\mathcal{L}_\infty$), which implies that $u\in\mathcal{L}_\infty$ in (\ref{eq:high u}) and $q,\dot{q}\in\mathcal{L}_\infty$. Furthermore, $\ddot{q}\in\mathcal{L}_\infty$.

\end{proof}
\end{theorem}

\begin{theorem}
\label{th:theorem2}
Given the closed-loop error system in (\ref{eq:low closed-loop}), the controller in (\ref{eq:low Ex u}) and (\ref{eq:low Fl u}) ensures exponential tracking in the sense that
\begin{equation}
    \Vert z_2(t)\Vert\leq\sqrt{\frac{b_{\varrho}}{a_{\varrho}}}\exp\left(-\frac{\lambda_{\varrho}}{2}(t-t_\omega)\right)\Vert z_2(t_\omega)\Vert,
    \label{eq:low exp result}
\end{equation}
$\forall t\in(0,\infty),\forall\omega$, where $\omega$ represents the $\omega^\text{th}$ time the system is activated and $\lambda_{\varrho}=\frac{1}{b_{\varrho}}\text{min}\{\beta,\underline{B}_n k_{2}\}$, provided the following sufficient gain conditions are satisfied
\begin{equation}
    k_{3}\geq\frac{c_1}{\underline{B}_n},k_{4}\geq\frac{c_2}{\underline{B}_n}.
    \label{cond:low exp cond}
\end{equation}
\begin{proof}
Defining a nonnegative, continuously differentiable Lyapunov function for the pair motor system $V_{\varrho}:\mathbb{R}\times\mathbb{R}\times\mathbb{R}_{\geq t_{0}}\rightarrow\mathbb{R}$ as
\begin{equation}
    V_{\varrho}=\frac{1}{2}e^2+\frac{1}{2}J_{\varrho}r^2,
    \label{eq:low V}
\end{equation}
which satisfies the following inequalities
\begin{equation}
    a_{\varrho}\left\| z_{2}\right\|^2\le V_{\varrho}(z_{2},t)\le b_{\varrho}\left\| z_{2}\right\|^2,
    \label{eq:low V bounded}
\end{equation}
where $a_\varrho=\min\left(\frac{1}{2},\frac{1}{2}c_j\right), b_\varrho=\max\left(\frac{1}{2},\frac{1}{2}c_J\right)$ are positive known constants. Since there exist discontinuous terms in motor synchronization control input (\ref{eq:low Ex u}) and (\ref{eq:low Fl u}), the Filippov method is used to analyze the system's stability. Let $z_{2}(t)$ be a Filippov solution to the differential inclusion $\dot{z}_2\in K[h](z_{2})$, where $K[\cdot]$ is defined as in \cite{Paden1987a} and $h$ is defined by using (\ref{eq:low closed-loop}) and (\ref{eq:low motor r}) as $h\triangleq[\begin{array}{cc}h_{1}&h_{2}\end{array}]$, where $h_{1}\triangleq r-\beta e$, $h_{2}\triangleq J_{\varrho}\dot{r}=\chi_{\varrho}-e-B_{\varrho}\left(k_{2} r+\left(k_{3}+k_{4}\left\Vert z_2\right\Vert\right)\text{sgn}(r)\right)$. The closed loop system in (\ref{eq:low closed-loop}) has discontinuous signum function, hence the time derivative of (\ref{eq:low V}) exists almost everywhere (a.e.), i.e., for almost all t. Based on \cite[Lemma 1]{Fischer2013}, the time derivative of (\ref{eq:low V}), $\dot{V}_{\varrho}(z_{2}(t),t)\overset{a.e.}{\in}\dot{\tilde{V}}_{\varrho}(z_{2}(t),t)$, where $\dot{\tilde{V}}_{\varrho}$ is the generalized time derivative of (\ref{eq:low V}) along the Filippov trajectories of $\dot{z}_2=h(z_{2})$ and is defined as in \cite{Fischer2013} as $\dot{\tilde{V}}_{\varrho}\triangleq\bigcap_{\xi\in\partial V_{\varrho}}\xi^{T}K\left[\begin{array}{ccc}\dot{e} & \dot{r} & 1\end{array}\right]^{T}(e,r,t)$. Since $V_{\varrho}(z_2,t)$ is continuously differentiable in $z_2$, $\partial V_{{\varrho}}=\{\nabla V_{{\varrho}}\}$, thus $\dot{\tilde{V}}_{{\varrho}}\overset{a.e.}{\subset}\left[\begin{array}{ccc}e&J_{\varrho}r&\frac{1}{2}\dot{J}_{\varrho}r^{2}\end{array}\right]K$ $\left[\begin{array}{ccc}
\dot{e} & J_{\varrho}\dot{r} & 1\end{array}\right]^{T}$. Therefore, after substituting for (\ref{eq:low closed-loop}) and canceling common terms, the generalized time derivative of (\ref{eq:low V}) can be expressed as
\begin{multline}
    \dot{\tilde{V}}_{\varrho}\overset{a.e.}{\subset}-\beta e^2+r\chi_{\varrho}
    \\-K[B_{\varrho}]r\big[k_{2} r+\left(k_{3}+k_{4}\left\Vert z_2\right\Vert\right)K[\text{sgn}(r)]\big],
    \label{eq:low V_dot}
\end{multline}
where $K[\text{sgn}(r)]=SGN(r)$ such that $SGN(r)=1$ if $r>0$; $[-1,1]$ if $r=0$; $-1$ if $r<0$, and $K[B_\varrho]\subset[\underline{B}_n,\bar{B}_n]$ as defined in~\cite{Fischer2013}.
With the condition in (\ref{eq:low chi upper bound}) and Property (\ref{prop:motor bounded_B}), (\ref{eq:low V_dot}) can be upper bounded as
\begin{multline}
    \dot{\tilde{V}}_{\varrho}\overset{a.e.}{\leq}-\beta e^2- \underline{B}_n k_{2} r^2 + \left(c_1-\underline{B}_n k_{3}\right)| r| 
    \\+ \left(c_2-\underline{B}_n k_{4}\right)| r|\Vert z_{2}\Vert.
    \label{eq:low V_dot bounded}
\end{multline}
Provided the sufficient gain conditions in (\ref{cond:low exp cond}), the following inequality is obtained
\begin{align}
    \dot{\tilde{V}}_{\varrho}&\overset{a.e.}{\leq}-\beta e^2- \underline{B}_n k_{2} r^2,
    \\&\overset{a.e.}{\leq}-\lambda_{\varrho}\tilde{V}_{\varrho}.
    \label{eq:low V_dot final}
\end{align}
Based on (\ref{eq:low V bounded}) and (\ref{eq:low V_dot final}), the exponential tracking result in (\ref{eq:low exp result}) can be obtained. Since $V_{\varrho}\in\mathcal{L}_\infty$, $e,r\in\mathcal{L}_\infty$ (i.e., $z_2\in\mathcal{L}_\infty$), which implies that $u_{ex},u_{fl}\in\mathcal{L}_\infty$ in (\ref{eq:low Ex u}) and (\ref{eq:low Fl u}).

\end{proof}
\end{theorem}
\begin{theorem}
\label{th:theorem3}
A paired motor system has an average dwell time $\tau_{a}$ if there exist positive numbers $N_{0}$ and $\tau_{a}$ such that
\begin{equation}
    N(T,t)\leq N_{0}+\frac{T-t}{\tau_{a}},\ \forall T\geq t\geq 0,
    \label{eq:MDADT number}
\end{equation}
where $N$ is the maximum allowable total switching. The switching motor system ensures exponential tracking with average dwell time \cite{Liberzon2003}
\begin{equation}
    \tau_{a}\geq\frac{\ln\left(\mu\right)}{\lambda_\varrho}.
    \label{eq:MDADT time}
\end{equation}
\begin{proof}
Let $t_{1},t_{2},...,t_{j},...,t_{N_{\sigma}}$ denotes the timing when the motor synchronization controller switched over the time interval $(0,T)$. The switching number for $\varrho=\{ex,fl\}$ motor system over the time interval $(0,T)$ is denoted as $N_{\varrho}(T,t)$, and the total switching number for the whole paired motor system can be written as $N(T,t)=N_{ex}(T,t)+N_{fl}(T,t)$. From the inequality in (\ref{eq:low V bounded}),
\begin{equation}
    V_{\varrho(t_{i+1})}\left(z_2(t_{i+1})\right)\leq \mu V_{\varrho(t_{i})}\left(z_2(t_{i+1})\right),
    \label{eq:V relation}
\end{equation}
where $\mu=\frac{b_\varrho}{a_\varrho}$. Define a piecewise differentiable function
\begin{equation}
    W(t)=e^{\lambda_\varrho t}V_{\varrho(t)}\left(z_2(t)\right),
    \label{eq:W function}
\end{equation}
which has a non-positive derivative equation
\begin{equation}
    \dot{W}(t)=\lambda_\varrho W(t)+e^{\lambda_\varrho t}\dot{V}_{\varrho(t)}\left(z_2(t)\right).
    \label{eq:W dot function}
\end{equation}
Defining $t_{i+1}^-$ the instantaneous time before $t_{i+1}$, so that the switching system $\varrho(t_{i+1}^-)=\varrho(t_{i+1})$. Considering the time at $t_{i+1}$ with the inequality in (\ref{eq:V relation}), non-increasing $W$ in (\ref{eq:W function}) can be expressed as
\begin{align*}
    W(t_{i+1})&\leq \exp\left[\lambda_\varrho t_{i+1}\right]\mu V_{\varrho(t_{i})}\left(z(t_{i+1})\right)
    \\&= \exp\left[\lambda_\varrho t_{i+1}\right]\mu V_{\varrho(t_{i+1}^{-})}\left(z(t_{i+1})\right)
    \\&= \mu W(t_{i+1}^{-}) \leq \mu W(t_{i}).
\end{align*}
Applying same method to $W(t_i)$ and the iteration yields the inequality relationship
\begin{equation}
    W(T^{-})\leq W(t_{N})\leq
    \\\mu^{N_{\sigma}(T,0)-1}W(0).
    \label{eq: W middle}
\end{equation}
Substituting (\ref{eq:W function}) into (\ref{eq: W middle}) yields
\begin{equation}
    \exp\left[\lambda_\varrho T\right]V_{\varrho(T^{-})}\left(z_2(T)\right)\leq\mu^{N(T,0)-1}V_{\varrho(0)}\left(z_2(0)\right).
    \label{eq: W middle 2}
\end{equation}
After performing algebraic manipulation and substituting the total switching number (\ref{eq:MDADT number}) into the equation yields
\begin{multline}
    V_{\varrho(T^{-})}\left(z_2(T)\right)\leq
    \\\exp\left[-\lambda_\varrho T+\left(N_{0}+\frac{T-t}{\tau_{a}}-1\right)\ln(\mu)\right]V_{\varrho(0)}\left(z_2(0)\right).
    \label{eq:MDADT middle}
\end{multline}
The inequality can be further bounded as
\begin{multline}
    V_{\varrho(T^{-})}\left(z_2(T)\right)\leq\exp\left[\left(N_{0}-1\right)\ln(\mu)\right]
    \\\exp\left[\left(\frac{\ln(\mu)}{\tau_a}-\lambda_\varrho\right)T\right]V_{\varrho(0)}\left(z_2(0)\right).
    \label{eq:MDADT exp}
\end{multline}
Given the condition in (\ref{eq:MDADT time}), $V_{\varrho(T^{-})}\left(z_2(T)\right)$ converges to zero exponentially.
\end{proof}
\end{theorem}
\section{Experiments}
An experiment is provided to demonstrate the performance of the controllers developed for kinematic joint tracking in (\ref{eq:high u}) and motor synchronization in (\ref{eq:low Ex u}) and (\ref{eq:low Fl u}) for leg swing. Results from one able-bodied individual (male aged 29 years) were obtained and written informed consent was obtained prior to participation as approved by the Institutional Review Board at Syracuse University. The participant was fitted with the exoskeleton, and instructed to keep the right leg standing while avoiding providing voluntary effort on the left leg.

The customized exoskeleton was designed for fitting a wide variety of body sizes and maintaining alignment with the user's joints. There were 2 brushless 24 VDC electric motors (Maxon) used to actuate the cable-driven mechanisms. The angles  of motor and joint were measured by Maxon motor's encoders and a optical encoder (US Digital), respectively. The controllers were implemented on a desktop computer (Windows 10 OS) running a real-time target (QUARC 2.6, Quanser) via MATLAB/Simulink 2018a (MathWorks Inc) with a sample rate of 1 kHz. A Quanser QPIDe and a Q-8 DAQ boards were used to read the encoders, and control the servo motor drivers (Maxon) operating in current-controlled mode.

\begin{figure}
    \centering
    \includegraphics[width=0.8\columnwidth]{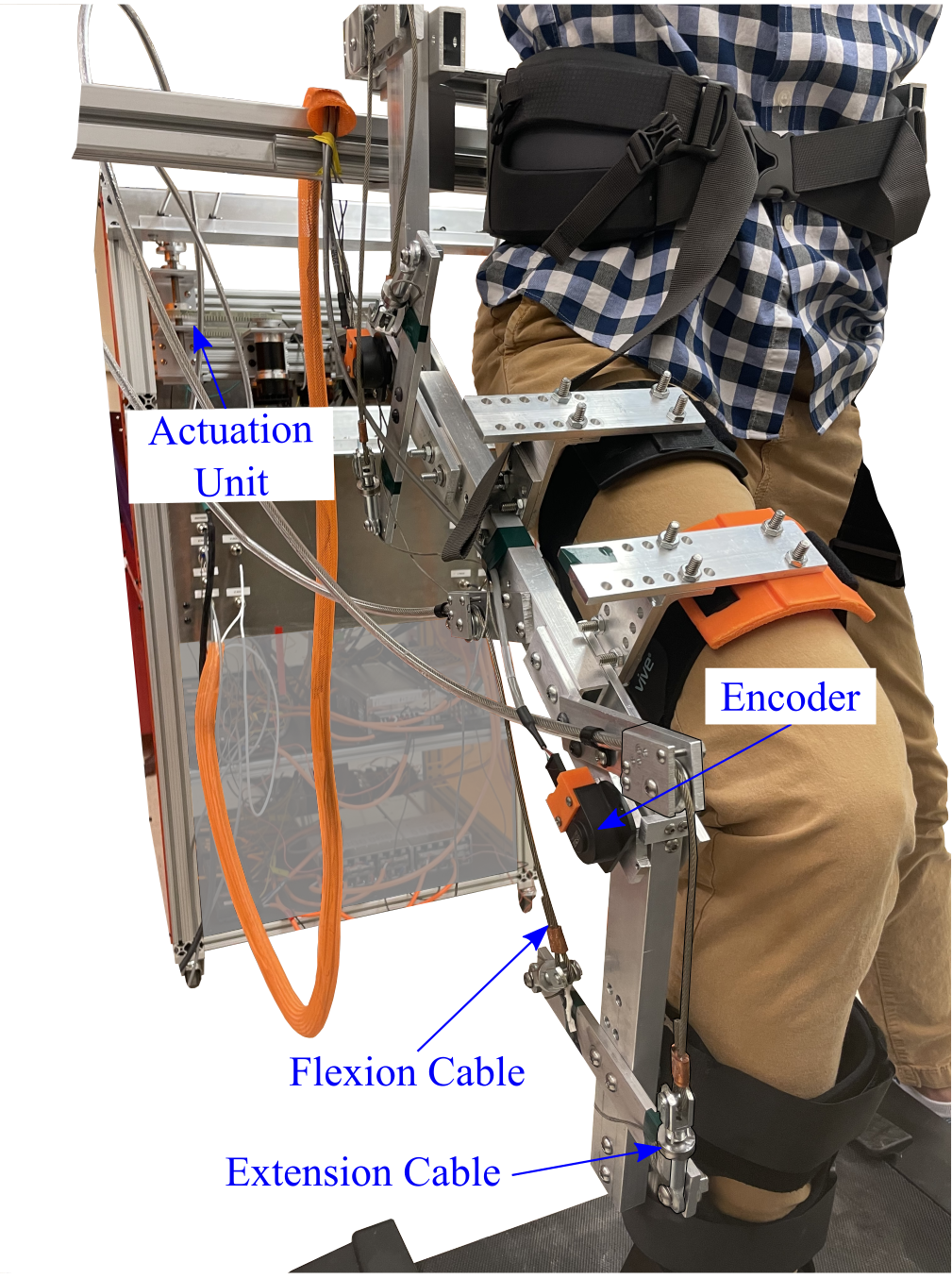}
    \caption{Cable-driven exoskeleton testbed used for experiment demonstration. The system uses flexion and extension cables to apply torque at each joint using electrical motors installed in the actuation unit.}
    \label{fig:exo}
\end{figure}

\begin{figure}
    \centering
    \begin{subfigure}
        \centering
        \includegraphics[width=0.8\columnwidth]{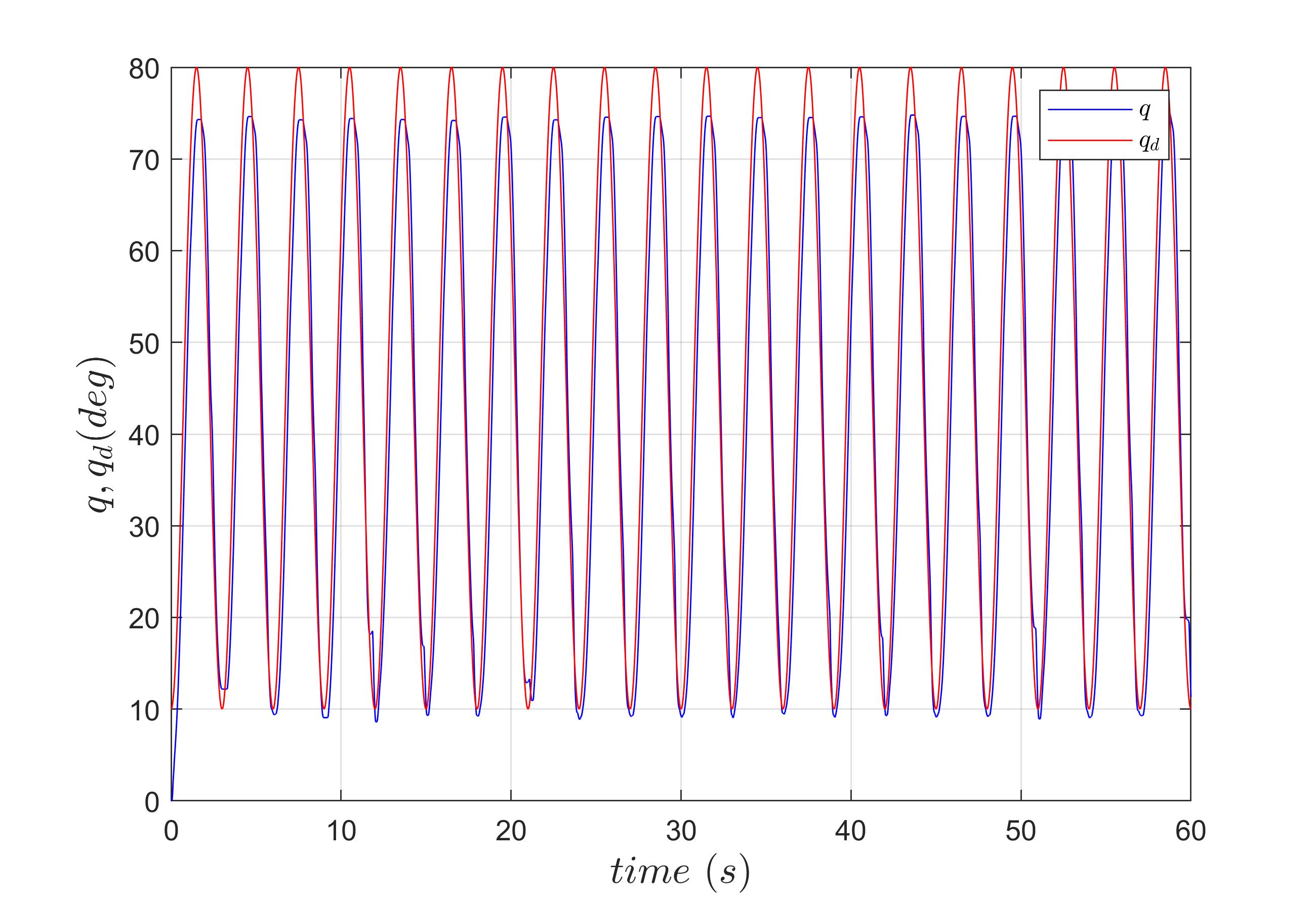}
        \caption{Joint angle trajectory tracking performance depicting the joint desired trajectory $q_d$ and actual joint trajectory $q$.}
        \label{fig:q_qd}
    \end{subfigure}
    \begin{subfigure}
        \centering
        \includegraphics[width=0.8\columnwidth]{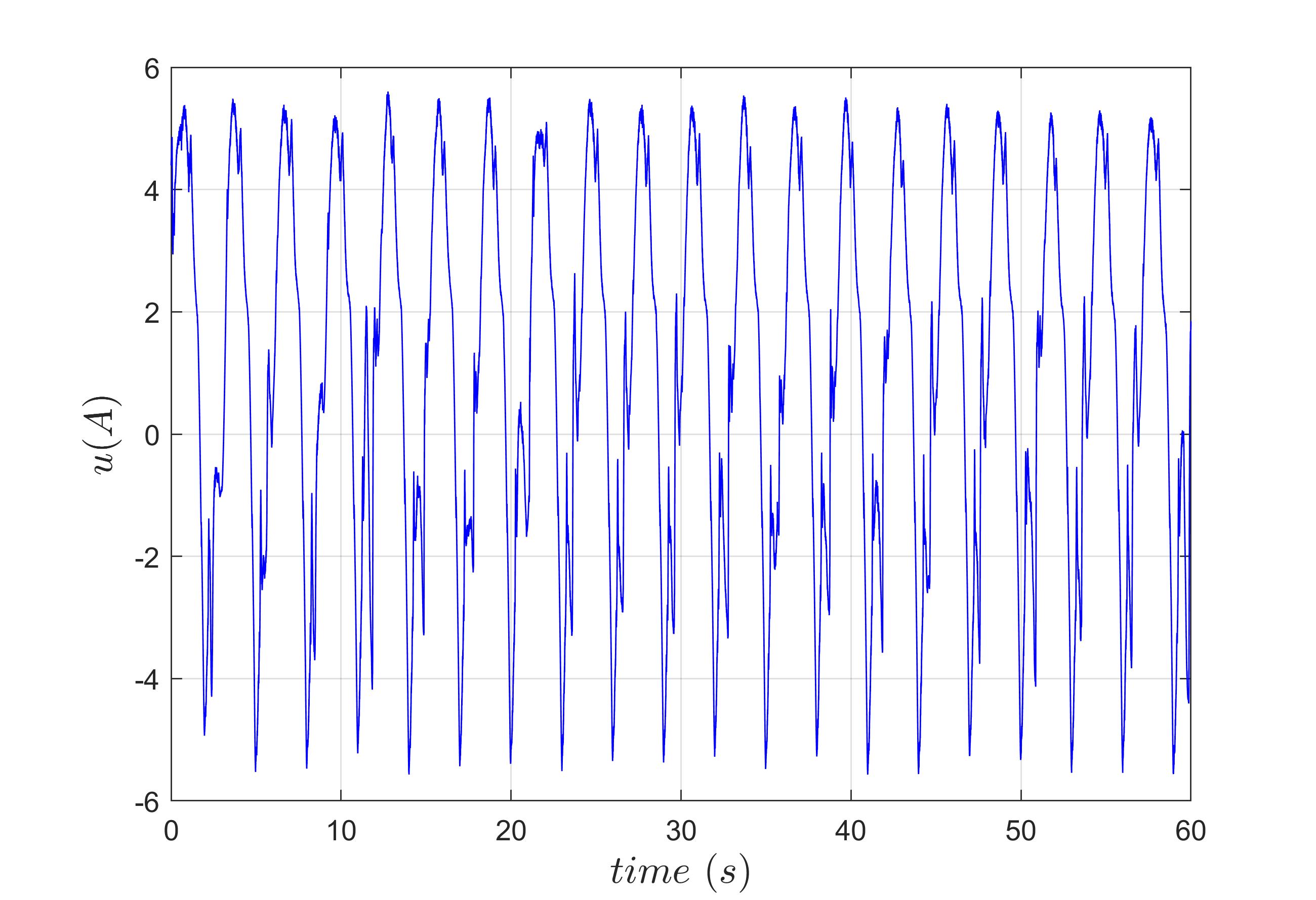}
        \caption{Control input $u$ for joint tracking.}
        \label{fig:u_exo}
    \end{subfigure}
    \begin{subfigure}
        \centering
        \includegraphics[width=0.8\columnwidth]{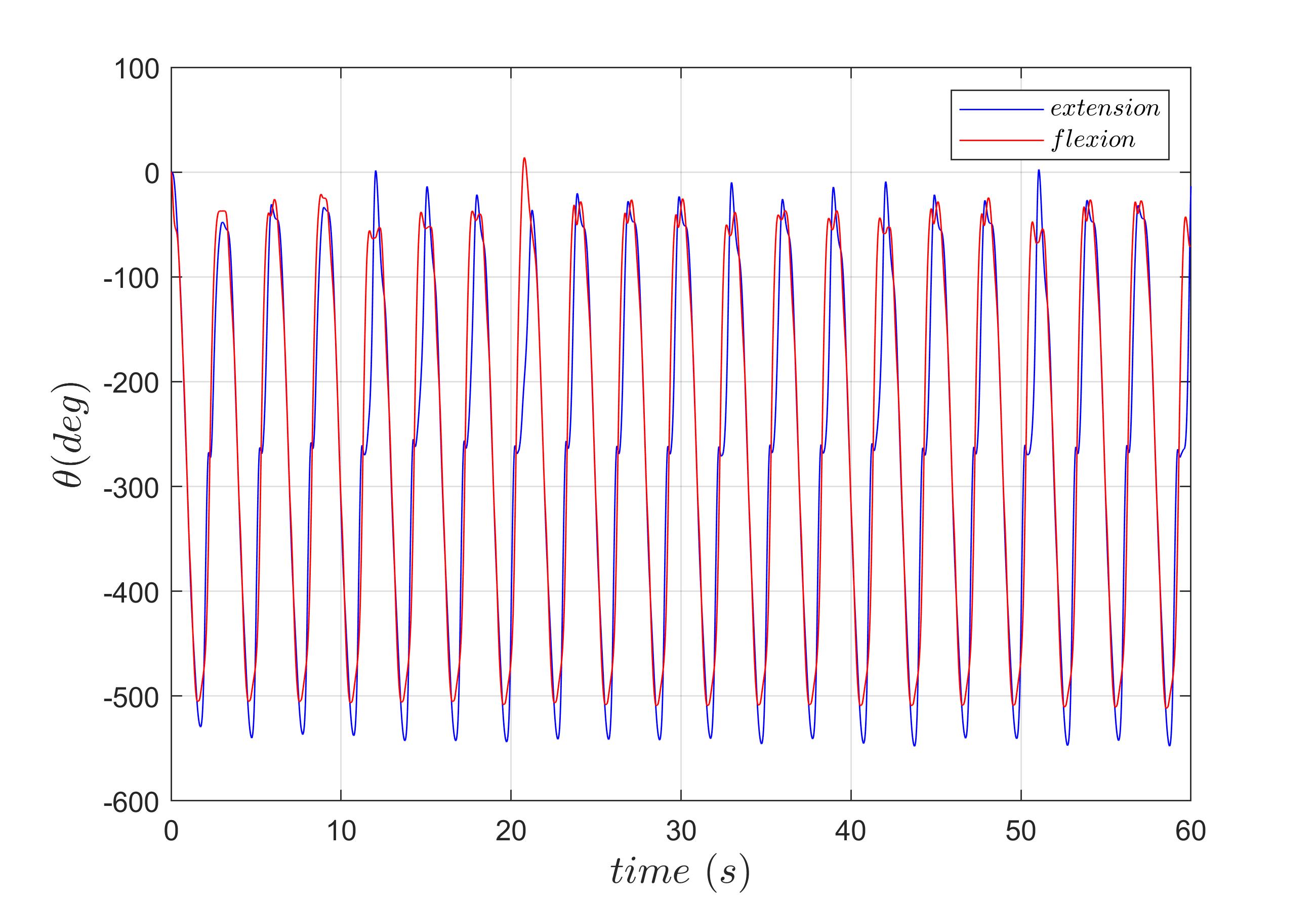}
        \caption{Motor synchronization trajectory tracking illustrating the motor flexion angle $\theta_{fl}$ and motor extension angle $\theta_{ex}$.}
        \label{fig:motor_angle}
    \end{subfigure}
    \begin{subfigure}
        \centering
        \includegraphics[width=0.8\columnwidth]{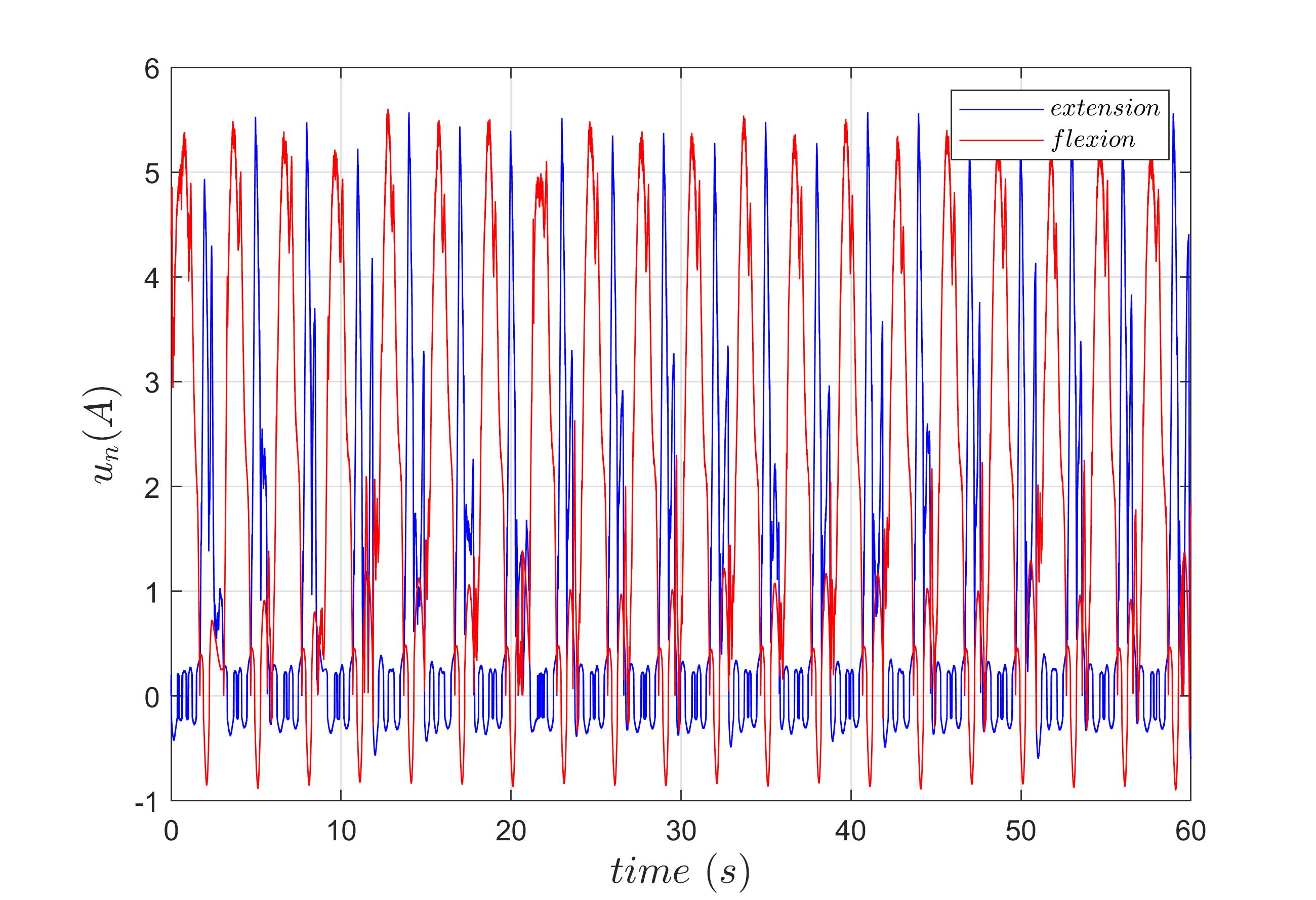}
        \caption{Synchronization control input for flexor $u_{fl}$ and extensor $u_{ex}$.}
        \label{fig:u_e}
\end{subfigure}
\end{figure}

Figure \ref{fig:exo} illustrates the exoskeleton testbed. The safety precautions include: an emergency stop button was installed near the user to stop the experiment manually, software stop conditions to ensure the testbed was performing in safe range of motion, and mechanical stops were designed on exoskeleton to avoid moving the legs through unsafe joint angles.
The desired left knee trajectory was designed to vary from 10 deg to 80 deg with a period of 3 seconds. The control gains introduced in (\ref{eq:high u}), (\ref{eq:low Ex u}) and (\ref{eq:low Fl u}) were selected as follows: $k_1=2, \epsilon=2, \alpha=10, k_2=0.01, k_3=0.2, k_4=0.0001, \beta=30$. The experiment was implemented for a duration of 60 seconds. Figure \ref{fig:q_qd} shows the desired and actual joint angles. The exoskeleton control input is presented in Figure \ref{fig:u_exo}. Figure \ref{fig:motor_angle} illustrates the performance of the synchronization tracking objective. The applied synchronization motor control inputs are depicted in Figure \ref{fig:u_e}.

\section{Conclusion}
The dynamical model for a human-exoskeleton and motor systems were introduced. A high-gain exoskeleton controller and a motor synchronization controller with sliding-mode method were developed. A Lyapunov-based stability analysis is developed to ensure globally ultimately bounded for the joint tracking objective, and exponential tracking for the motor synchronization objective. The average dwell time condition provides a bound on the number of motor switches and guarantees exponential tracking for the switching motors. Future work includes developing a force controller to improve tracking performance, and evaluate the control performance in walking experiments.





\AtNextBibliography{\small}
\printbibliography
\end{document}